\documentclass{aa}

\usepackage{amsmath}
\usepackage{amssymb}
\usepackage{upgreek}
\usepackage[varg]{txfonts}
\usepackage{epsfig,graphicx}
\usepackage{color}
\usepackage[colorlinks=true, linkcolor=blue, citecolor=blue, urlcolor=blue]{hyperref}
\usepackage{natbib}
\bibpunct{(}{)}{;}{a}{}{,} 
\defcitealias{2015arXiv151103789S}{SS15}

\begin{document}

\title{The nature of the UV halo around the spiral galaxy NGC\,3628}
\titlerunning{The UV halo around NGC\,3628}
\author{Maarten Baes \and S\'ebastien Viaene}
\authorrunning{M. Baes \& S. Viaene}
\institute{Sterrenkundig Observatorium, Universiteit Gent, Krijgslaan 281 S9, B-9000 Gent, Belgium}

\abstract{%
Thanks to deep UV observations with {\it{GALEX}} and {\it{Swift}}, diffuse UV haloes have recently been discovered around galaxies. Based on UV-optical colours, it has been advocated that the UV haloes around spiral galaxies are due to UV radiation emitted from the disc and scattered off dust grains at high latitudes. Detailed UV radiative transfer models  that take into account scattering and absorption can explain the morphology of the UV haloes, and they require the presence of an additional thick dust disc next the to traditional thin disc for half of the galaxies in their sample. We test whether such an additional thick dust disc agrees with the observed infrared emission in NGC\,3628, an edge-on galaxy with a clear signature of a thick dust disc. We extend the far-ultraviolet radiative transfer models to full-scale panchromatic models. Our model, which contains no fine-tuning, can almost perfectly reproduce the observed spectral energy distribution from UV to mm wavelengths. These results corroborate the interpretation of the extended UV emission in NGC\,3628 as scattering off dust grains, and hence of the presence of a substantial amount of diffuse extra-planar dust. A significant caveat, however, is the geometrical simplicity and non-uniqueness of our model: other models with a different geometrical setting could lead to a similar spectral energy distribution. More detailed radiative transfer simulations that compare the model results to images from UV to submm wavelengths are a way to break this degeneracy, as are UV polarisation measurements.
}


\maketitle

\section{Introduction}

The amount and spatial distribution of dust in spiral galaxies has been a hot topic for several decades. One of the most interesting possibilities is that substantial amounts of interstellar dust could reside in the haloes of spiral galaxies. The first tentative detection of dust in the haloes of spiral galaxies was made by \citet{1994AJ....108.1619Z}, based on systematic reddening of background galaxies behind two nearby spiral galaxies \citep[see also][]{2006ApJ...651L.107X}. An impressive effort was made by \citet{2010MNRAS.405.1025M}, who used SDSS colours of 85,000 quasars to statistically estimate the reddening by dust extinction at large distances from foreground galaxies. They claimed the detection of systematic dust reddening out to several Mpc and argued that half of all the interstellar dust in the Universe should reside in the haloes of normal spiral galaxies. Consistent conclusions were reached by \citet{2015ApJ...813....7P} based on a similar analysis with passively evolving galaxies as background "standard crayons".

Several recent studies have attempted to detect the presence of dust in the haloes of individual galaxies. Extra-planar dust has been detected in extinction in optical bands in the form of filamentary structures \citep{1997AJ....114.2463H, 1999AJ....117.2077H, 2004AJ....128..662T}. These features can be detected up to limited distances above the plane of a galaxy, and their contribution to the total dust mass budget in a galaxy is rather limited. This suggests that a larger diffuse component should be present that is hard to detect using extinction at optical wavelengths. One method to trace such a diffuse component is by searching for thermal dust emission \citep{2007ApJ...668..918B, 2009ApJ...698L.125K, 2013A&A...556A..54V, 2015arXiv150907677B}. Complementary to diffuse dust emission, various teams have detected extra-planar polycyclic
aromatic hydrocarbon (PAH) emission in nearby edge-on galaxies \citep{2006A&A...445..123I, 2007A&A...474..461I, 2009MNRAS.395...97W, 2013ApJ...774..126M}. 

An interesting alternative method to search for dust in the haloes of galaxies is by using UV observations. UV radiation is scattered very efficiently by interstellar dust, and the sky is extremely dark at these wavelengths, such that even very faint surface brightness levels can be detected. These facts stimulated \citet{2014ApJ...789..131H} to search for the presence of diffuse UV haloes around nearby galaxies, based on {\em{GALEX}} and {\em{Swift}} images. They found clear signatures of diffuse UV emission around late-type galaxies out to 20 kpc from the galaxy mid-planes. Based on the colours, they argued that this diffuse emission is most likely due to starlight that has escaped from the galaxies and scattered off dust grains in the halo. 

The conclusion that the diffuse UV emission around edge-on spiral galaxies is due to dust scattering was  corroborated by \citet[][hereafter \citetalias{2015arXiv151103789S}]{2015arXiv151103789S}, who extended a pilot study of \citet{2014ApJ...785L..18S}. They selected six galaxies from the sample of \citet{2014ApJ...789..131H} that seemed reasonably well described by exponential distributions in the vertical and radial directions. They fitted an axisymmetric galaxy model to the {\it{GALEX}} far-ultraviolet (FUV) images of each of these six galaxies by means of a Monte Carlo radiative transfer code. The dust in their model was distributed in two layers: a thick and a thin one, each with its own scale height and optical depth. Interestingly, \citetalias{2015arXiv151103789S} found that three galaxies in their sample showed clear evidence for a thick dust layer, whereas for the remaining three galaxies the vertical surface brightness profile could be modelled satisfactorily with just a single thin dust disk. 

A critical point in the analysis and interpretation of \citetalias{2015arXiv151103789S} is that they performed their radiative transfer modelling in a single waveband. They determined the amount and spatial distribution of the dust in their models by only using the UV attenuation, but it is not necessarily guaranteed that their models also agree with other constraints on the dust content in galaxies. A powerful way to test the validity of their models is to consider the infrared emission that they would predict. In particular, we could critically investigate whether the amount and spatial distribution of dust that is necessary to explain the UV emission from the halo agrees with infrared and submm flux measurements \citep[see also][]{2010A&A...518L..39B, 2012MNRAS.427.2797D, 2012MNRAS.419..895D, 2015MNRAS.451.1728D, 2015A&A...579A.103V}.

In this paper, we test the claims of \citetalias{2015arXiv151103789S} about the nature of the UV halo of NGC\,3628 by expanding their monochromatic FUV radiative transfer model to a panchromatic model that covers the entire wavelength range from UV to mm wavelengths. We select this galaxy because it is the most extreme case in their study: of the three galaxies with a claimed clear detection of a thick dust layer, it is the most massive galaxy with the highest star formation rate. According to the modelling, the thick disc of NGC\,3628 has a surprisingly large optical depth, and accordingly, a significant fraction of the dust mass was found to be located  outside the main disk. It is therefore an ideal galaxy to test whether the UV-based attenuation model agrees with the observed infrared data.

In Sect.~{\ref{Data.sec}} we present the panchromatic data set we use for our modelling. In Sect.~{\ref{SKIRT.sec}} we present our panchromatic radiative transfer model and in particular the predictions in the infrared--submm region, and compare this to the observational data. In Sect.~{\ref{Discussion.sec}} we discuss the results of our modelling and present our conclusions.

\section{Data}
\label{Data.sec}

\begin{table*}
\caption{Integrated broad-band flux densities for NGC\,3628.}
\centering
\begin{tabular}{cccl}
\hline\hline \\ 
band & $\lambda$ [$\upmu$m] & $F_\nu$ [Jy] & source \\
\\ \hline \\
GALEX FUV & 0.153 & $0.00738\pm0.00061$ & \citet{2011ApJS..192....6L} \\
GALEX NUV & 0.227 & $0.0194\pm0.0005$ & \citet{2011ApJS..192....6L} \\
SDSS {\it{u}} & 0.358 & $0.105\pm0.005$ & this work \\
Johnson B & 0.439 & $0.329\pm0.016$ & \citet{1991rc3..book.....D} \\
SDSS {\it{g}} & 0.485 & $0.419\pm0.021$ & this work \\
Johnson V & 0.552 & $0.588\pm0.040$ & \citet{1991rc3..book.....D} \\ 
SDSS {\it{r}} & 0.627 & $0.797\pm0.040$ & this work \\
SDSS {\it{i}} & 0.769 & $1.170\pm0.059$ & this work \\
SDSS {\it{z}} & 0.912 & $1.531\pm0.077$ & this work \\ 
2MASS {\it{J}} & 1.25 & $2.15\pm0.040$ & \citet{2003AJ....125..525J} \\
2MASS {\it{H}} & 1.64 & $2.88\pm0.062$ & \citet{2003AJ....125..525J} \\
2MASS {\it{K}}$_{\text{s}}$ & 2.17 & $2.48\pm0.053$ & \citet{2003AJ....125..525J} \\
WISE W1 & 3.37 & $1.509\pm0.037$ & this work \\
IRAC 3.6 & 3.55 & $1.52\pm0.21$ & \citet{2009ApJ...703..517D} \\
IRAC 4.5 & 4.49 & $1.04\pm0.14$ & \citet{2009ApJ...703..517D} \\
WISE W2 & 4.62 & $0.928\pm0.027$ & this work \\
IRAC 5.8 & 5.72 & $1.86\pm0.23$ & \citet{2009ApJ...703..517D} \\
IRAC 8.0 & 7.85 & $4.08\pm0.51$ & \citet{2009ApJ...703..517D} \\
IRAS 12 & 12.0 & $3.13\pm0.05$ & \citet{2003AJ....126.1607S} \\
WISE W3 & 12.1 & $4.812\pm0.223$ & this work \\
WISE W4 & 22.2 & $4.946\pm0.305$ & this work \\
MIPS 24 & 23.6 & $5.10\pm0.55$ & \citet{2009ApJ...703..517D} \\
IRAS 25 & 24.9 & $4.85\pm0.063$ & \citet{2003AJ....126.1607S} \\ 
IRAS 60 & 59.9 & $54.8\pm0.077$ & \citet{2003AJ....126.1607S} \\
MIPS 70 & 71.3 & $68.6\pm8.40$ & \citet{2009ApJ...703..517D} \\
IRAS 100 & 99.8 & $106\pm0.24$ & \citet{2003AJ....126.1607S} \\
MIPS 160 & 156 & $190\pm30$ & \citet{2009ApJ...703..517D} \\
Planck 857 & 350 & $43.57\pm0.84$ & \citet{2015arXiv150702058P} \\
SCUBA 450 & 449 & $16.9\pm0.196$ & \citet{2005MNRAS.357..361S} \\
Planck 545 & 550 & $12.27\pm0.36$ & \citet{2015arXiv150702058P} \\
SCUBA 850 & 848 &       $2.82\pm0.18$ & \citet{2005MNRAS.357..361S} \\
Planck 353 & 849 & $3.120\pm0.166$ & \citet{2015arXiv150702058P} \\
Planck 217 & 1382 &      $0.569\pm0.100$ & \citet{2015arXiv150702058P} \\
\\ \hline
\end{tabular}
\end{table*}

NGC\,3628 is a well-known edge-on Sb galaxy at a distance of 10.95~Mpc.\footnote{We adopt the same distance to NGC\,3628 as \citetalias{2015arXiv151103789S}, who used the mean of a suite of different 
redshift-independent distance estimates from the literature, mainly based on the Tully-Fisher relation.} It forms the so-called Leo Triplet together with NGC\,3623 and NGC\,3627. In the optical, the galaxy is heavily obscured by a prominent dust lane. It hosts a strong nuclear starburst and has been studied extensively in almost all wavelength regimes, including X-rays \citep{1996ApJ...461..724D, 2004ApJS..151..193S}, submillimeter \citep{2005MNRAS.357..361S}, CO lines \citep{2009A&A...506..689I, 2012ApJ...752...38T}, and radio continuum \citep{2013A&A...553A...4N}.

For our purposes, we need a well-sampled spectral energy distribution covering the UV to mm wavelength range. In the ultraviolet, we used the flux densities from the GALEX Local Volume Legacy survey \citep{2011ApJS..192....6L}. At optical wavelengths, {\it{B}} and {\it{V}} flux densities were taken from the RC3 catalogue \citep{1991rc3..book.....D}, and {\it{urgiz}} fluxes were determined using aperture photometry from the Sloan Digital Sky Survey DR12 images \citep{2015ApJS..219...12A}. In the infrared, we used {\em{JHK}} flux densities from the 2MASS Large Galaxy Atlas \citep{2003AJ....125..525J}, {\it{Spitzer}} IRAC and MIPS flux densities from the Spitzer Local Volume Legacy Survey \citep{2009ApJ...703..517D}, IRAS flux densities from the IRAS Revised Bright Galaxy Sample \citep{2003AJ....126.1607S}, and WISE flux densities determined from the images following the procedures described by \citet{2015MNRAS.451.3815A}. Finally, in the submm--mm range, we used the SCUBA flux densities from \citet{2005MNRAS.357..361S} and the Planck data from the Second Planck Catalog of Compact Sources \citep{2015arXiv150702058P}. In total, we have 33 data points between 0.15~$\upmu$m and 1.4~mm. 

\section{Radiative transfer modeling}
\label{SKIRT.sec}

The goal of this paper is to investigate whether the FUV monochromatic radiative transfer model of \citetalias{2015arXiv151103789S} for NGC\,3628 is compatible with the observed spectral energy distribution from UV to mm wavelengths. More specifically, we wish to test whether the amount and spatial distribution obtained by \citetalias{2015arXiv151103789S} from FUV attenuation modelling gives rise to a dust emission spectrum that can reproduce the observed infrared flux densities. 

We used the 3D radiative transfer code SKIRT \citep{2003MNRAS.343.1081B, 2011ApJS..196...22B, 2015A&C.....9...20C} for this purpose. SKIRT is a versatile and flexible dust radiative transfer code based on the Monte Carlo technique. With its extended suite of input models for the stars and the dust \citep{2015A&C....12...33B} and advanced grid structures to partition the configuration space \citep{2013A&A...554A..10S, 2014A&A...561A..77S, 2013A&A...560A..35C}, it is designed to solve complex dust radiative transfer problems in an arbitrary 3D geometry. Crucial for our purposes is that dust emission is included in a self-consistent way. SKIRT can handle both equilibrium and transient dust emission from arbitrary dust mixtures, using either a brute force or a library approach \citep{2011ApJS..196...22B, 2015A&A...580A..87C}.

For our panchromatic radiative transfer modelling, we need to specify the spatial and spectral properties of the sources (the stars), the spatial distribution and properties of the dust, and the position of the observer with respect to the galaxy. The last point can directly be reproduced from \citetalias{2015arXiv151103789S}. The galaxy is at a distance of 10.95~Mpc from the observer and is seen under an inclination of 88.4 degrees, that is, 1.6 degrees from exactly edge-on. 

\begin{figure*}
\centering
\includegraphics[width=0.9\textwidth]{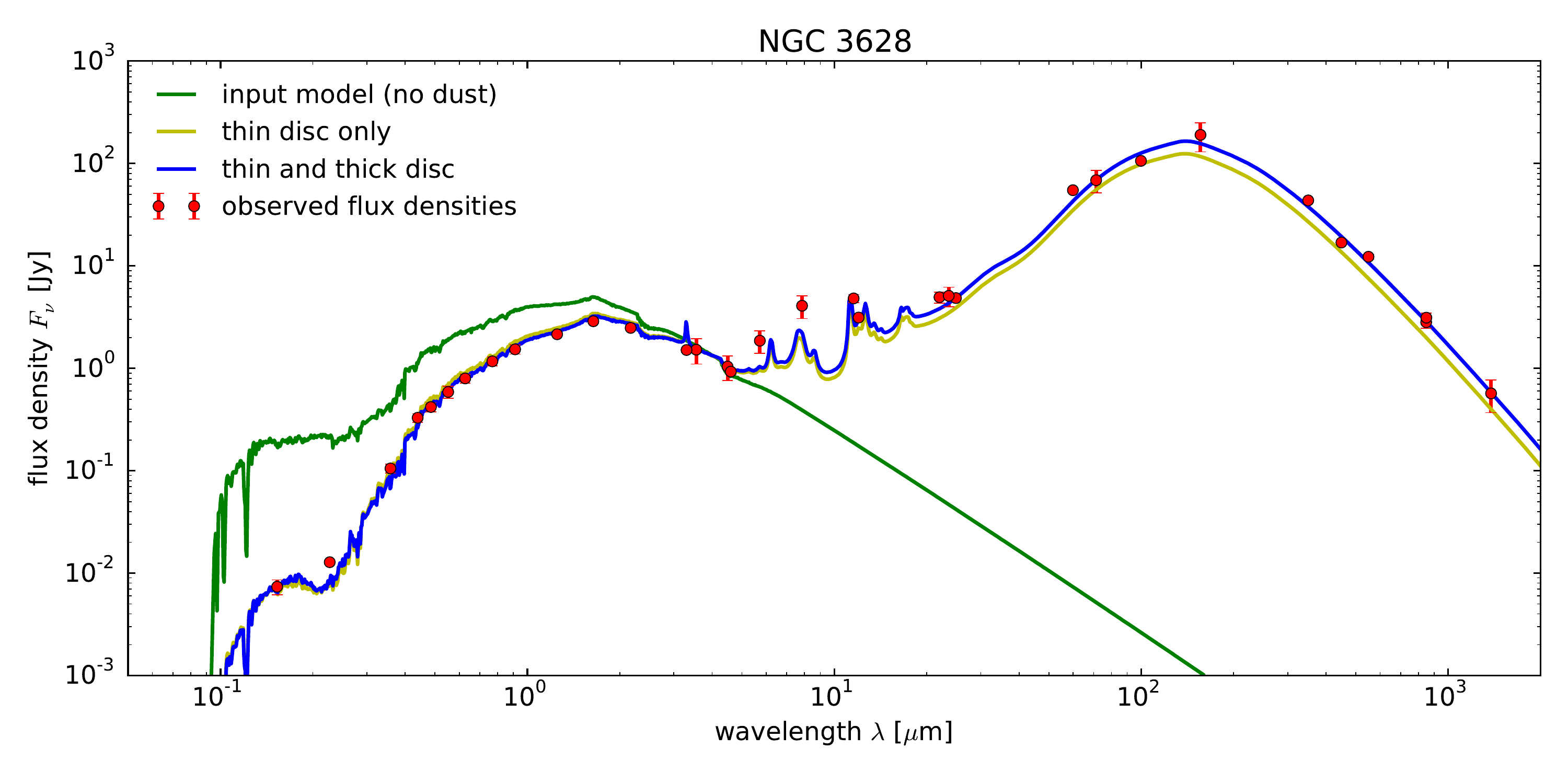}
\caption{UV--mm spectral energy distribution of NGC\,3628. The red data points are collected from the literature, as discussed in Sect.~{\ref{Data.sec}}. The solid green line is the input SED model, which consists of an evolved (5 Gyr) stellar population normalised to the IRAC~3.6~$\upmu$m flux density, and a young (50 Myr) stellar population normalised to the {\it{GALEX}} FUV flux density. The solid blue line is the SED from our panchromatic radiative transfer model, with a thin and thick disc. The solid yellow line is the SED corresponding to a model with only a thin disc; it systematically underestimates the observed flux densities in the far-infrared and submm range.}
\label{sed.fig}
\end{figure*}

For the spatial distribution of the dust, we used the model predictions by \citetalias{2015arXiv151103789S}, meaning that we assume two exponential discs. Both discs share the same scale length of 16.4~kpc and a truncation of 20~kpc. The thin disc has a scale height of 225~pc, whereas the thick disc has a scale height of 2.1~kpc. The B-band face-on optical depths of the two discs are 1.23 and 0.64, respectively.  This corresponds to a total dust mass of $1.30\times10^8~M_\odot$, with the thin disc contributing 66\% and the thick disc contributing the remaining 34\%. In our modelling, we used the BARE\_GR\_S dust model from \citet{2004ApJS..152..211Z}, which is designed to reproduce the FUV to near-infrared (NIR)
 extinction curve, the infrared emission, and depletion of the diffuse interstellar medium in the Milky Way.

We cannot directly extrapolate the model results of \citetalias{2015arXiv151103789S}
for the sources because they only refer to the FUV band, and we need a panchromatic input model. We made the very simple assumption that the stars in NGC\,3628 can be modelled as two distinct components: a young, star-bursting component, and an evolved component. For the former, we adopted the geometry from the \citetalias{2015arXiv151103789S} model, that is, a thin exponential disc with scale length of 7.8~kpc, radial truncation of 19.5~kpc, and scale height of 164~pc. We assumed a 50 Myr old SSP from the library of \citet{2005MNRAS.362..799M}. For the evolved stellar component, we used the geometry that has been fit to the IRAC 3.6~$\upmu$m image of NGC\,3628 by \citet{2015ApJS..219....4S} in the frame of the S$^4$G project \citep{2010PASP..122.1397S}. It consists of a combination of two edge-on discs with an exponential distribution in the radial and an isothermal sech$^2$ distribution in the vertical direction. We also assumed a \citet{2005MNRAS.362..799M} SSP template here, this time with an age of 5~Gyr. 

The only two parameters that remain to be set are the luminosities of each of the stellar components. This is relatively easy since they each dominate a completely different wavelength range. The luminosity of the young stellar component was determined such that it reproduced the observed FUV flux. This was done by running a radiative transfer simulation with a single component with a dummy luminosity, calculating the model flux density in the FUV band, and determining the scaling factor by comparing this to the observed FUV flux density. The FUV luminosity of the young component was found to be $\nu L_\nu = 1.43\times10^{10}~L_\odot$. Using the relation between star formation rate and FUV luminosity presented by \citet{2012ARA&A..50..531K}, this corresponds to a star formation rate (SFR) of $2.45~M_\odot\,{\text{yr}}^{-1}$, which perfectly agrees with the results of \citetalias{2015arXiv151103789S}. A similar approach was followed to determine the luminosity of the evolved stellar component, but now based on the IRAC 3.6~$\upmu$m flux density. The resulting luminosity is $\nu L_\nu = 4.76\times10^9~L_\odot$. Following the relations of \citet{2015ApJS..219....5Q}, this corresponds to a total stellar mass of $5.6\times10^{10}~M_\odot$.

With all these parameter fixed, we ran a panchromatic SKIRT simulation covering the entire wavelength range from UV to mm wavelengths, and in particular, we predicted the infrared spectral energy distribution of NGC\,3628. The resulting spectral energy distribution (SED) is shown in Fig.~{\ref{sed.fig}}, together with the observed flux densities. The result is astonishing: the model reproduces the data remarkably well over the entire wavelength range. We note again that the solid blue line {\em{\textup{is not a fit to the data points}}}: the model is completely determined by the FUV attenuation model of \citetalias{2015arXiv151103789S} and the flux densities in the FUV and IRAC 3.6~$\upmu$m bands. No constraints are imposed in the infrared domain, and there has been no fine-tuning at all.

\section{Discussion}
\label{Discussion.sec}

\subsection{Implications of the results}

Our modelling results demonstrate that the radiative transfer model of \citetalias{2015arXiv151103789S} agrees excellently
well with the UV--mm SED. In particular, this model, which was only based on the attenuation in the FUV bands, predicts the dust emission in the infrared--mm wavelength range almost perfectly. 

We note that it is definitely not obvious or expected a priori that this "blind" panchromatic modelling returns an SED that agrees so well with the observations. In the recent past, there have been several studies in which the amount and spatial distribution of dust in edge-on spiral galaxies has been estimated using radiative transfer modelling of optical images.  This panchromatic model has subsequently been used in these studies to predict the infrared--mm emission. Most of these so-called energy balance studies found a clear deficit in the energy balance, in the sense that the optically based radiative transfer models underestimated the observed infrared flux densities by a factor of about three \citep[e.g.][]{2000A&A...362..138P, 2011A&A...527A.109P, 2001A&A...372..775M, 2010A&A...518L..39B, 2012MNRAS.427.2797D, 2012MNRAS.419..895D}. This inconsistency has become known as the dust energy balance problem. 

A recent study by \citet{2015MNRAS.451.1728D} demonstrated that this dust energy balance inconsistency is not always present, and if it is present, not always to the same degree. They modelled two edge-on spiral galaxies for which a similar panchromatic data set was available using the same oligochromatic radiative transfer fitting technique \citep{2014MNRAS.441..869D}. For one galaxy, they found that the radiative transfer model underestimated the observed infrared fluxes by a factor of about three, whereas for the other galaxy, both the predicted SED and the simulated images matched the observations particularly well. This study makes it unlikely that one single physical mechanism is responsible for the dust energy balance problem in spiral galaxies. Several scenarios that could contribute to this problem have been proposed, including an underestimation of the FIR/submm emissivity \citep{2004A&A...425..109A, 2005A&A...437..447D}, the existence of an additional thin and dense dust disc \citep{2000A&A...362..138P, 2007MNRAS.379.1022D}, or the presence of large- and small-scales structure and inhomogeneities in the interstellar medium \citep{2008A&A...490..461B, 2010A&A...518L..39B, 2014A&A...571A..69D, 2015A&A...576A..31S}. 

To investigate the importance of the thick disc component, we reran our simulations without it (i.e., with only the thick dust disc). The resulting SED is shown as the yellow line in Fig.~{\ref{sed.fig}}. This SED systematically underestimates the observed flux densities at infrared to mm wavelengths, in a similar way as in many of the edge-on galaxies that have been modelled with single dust discs. These results corroborate the interpretation of the extended UV emission in NGC\,3628 as scattering off dust grains, and hence of the presence of a substantial amount of diffuse extra-planar dust. 

More generally, the present study, in combination with the previous work by \citet{2014ApJ...789..131H}, \citet{2014ApJ...785L..18S} and \citetalias{2015arXiv151103789S}, suggests that a vertically extended dust distribution might also be a viable contribution for the dust energy balance problem, at least for some galaxies. 

The possible existence of such diffuse dusty haloes would also have other serious implications. If dust in galactic haloes is a widespread phenomenon, it might affect the colours of background galaxies \citep{1994AJ....108.1619Z, 2006ApJ...651L.107X, 2010MNRAS.405.1025M}. This would imply that a significant fraction of the gas in the galactic haloes could be in a cold phase, which would agree with recent predictions from numerical simulations. Using hydrodynamical simulations, \citet{2013ApJ...770..139C} predicted that cold gas forms the dominant phase in the inner $\sim$150 kpc around spiral galaxies, while hot gas only starts to dominate from distances larger than about 200~kpc. It also raises questions on the evolution of the interstellar medium and the interaction between galaxies and their haloes. Several mechanisms can contribute to the expulsion of dust grains into the galactic halo, including radiation pressure, magnetic field effects, and hydrodynamical instabilities \citep[][and references therein]{1991ApJ...381..137F, 1998MNRAS.300.1006D, 1997AJ....114.2463H}. NGC\,3628 is a starburst galaxy and is known to have a central outflow of ionised and molecular gas \citep{2004ApJ...606..829S, 2012ApJ...752...38T}. Such an outflow seems the most probable mechanism to lift dust grains out of the central plane of a galaxy.

\subsection{Caveats and future investigation}

Before finally concluding and firmly stating that the nature of UV haloes around spiral galaxies is uncovered, it is necessary to critically investigate possible caveats in the analysis and alternative explanations.

\begin{figure}
\centering
\includegraphics[width=\columnwidth]{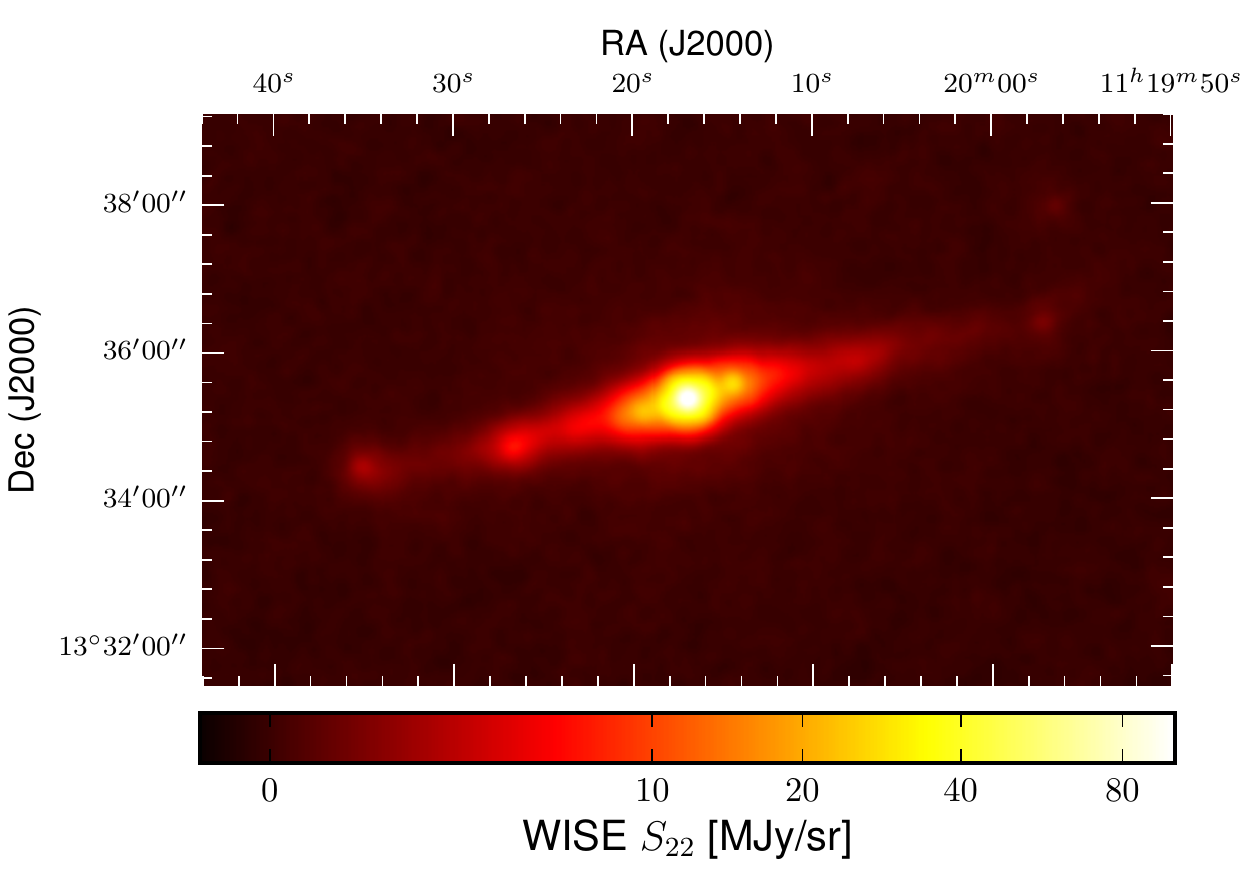}
\caption{WISE 22~$\upmu$m image of NGC\,3628. The surface brightness is plotted on a logarithmic scale.}
\label{NGC3628-WISE.fig}
\end{figure}

An important and possibly crucial caveat in our modelling is the simplicity of the geometry of our model. In the \citetalias{2015arXiv151103789S} model, most of the UV emission and hence the star formation is located in a thin and elongated disc, and the dust is distributed in two smooth and extended discs. As a result, the far-infrared emission in our panchromatic model is smoothly distributed over the entire galaxy. However, NGC\,3628 is characterised by a central nuclear starburst with a diameter of less than 1 kpc that may account for half of the total star formation in the galaxy \citep{1996ApJ...464..738I, 2015PKAS...30..499T}. Figure~{\ref{NGC3628-WISE.fig}} shows the WISE 22~$\upmu$m image of NGC\,3628, which mainly traces the emission from warm dust in star-forming regions \citep{2012JApA...33..213S, 2013AJ....145....6J, 2013ApJ...774...62L}. The surface brightness distribution is clearly peaked towards the centre. The fact that our model with a thick dust layer reproduces the observed SED could hence be misleading: other models with a different geometrical setting, in particular with a central concentration of dust rather than a vertically extended distribution, could lead to a similar infrared SED.

Our analysis indicates that a more extended analysis is required to break this degeneracy. In particular, it would be useful to run a suite of radiative transfer simulations with a sufficiently large variation in star--dust geometry and compare them to images across the UV--submm wavelength range instead of to the observed SED alone. A major difficulty here is the poor angular resolution of the available infrared instrumentation and the crucial dependence on an accurate knowledge of the shape of the point spread function \citep{2015arXiv150907677B}. Such an extended study is beyond the scope of this paper.

In spite of the simplicity of our model and the fact that it probably is a rather poor representation of the true intrinsic structure of NGC\,3628, the observed distribution of FUV emission remains a remarkable feature of this galaxy. The only serious contender for the thick dust disc scenario is that the UV emission is due to direct emission of a population of halo stars. Based on UV--optical colours, \citet{2014ApJ...789..131H} argued that direct stellar emission is likely responsible for the UV emission in the haloes around early-type galaxies, but that scattered emission of radiation escaped from the main disc is more likely for late-type galaxies. 

A way to test the nature of the diffuse UV emission in NGC\,3628 that is completely independent from the panchromatic radiative transfer analysis that we have presented is UV polarimetry, as also suggested by \citet{2014ApJ...789..131H}. If the UV halo emission is dominated by scattered radiation, we would expect a strong linear polarisation signature, similar as seen in reflection nebulae. Unfortunately, there is currently no UV polarisation instrument available. An option could be to move to slightly longer wavelengths, for example in the U band, where efficient polarimetric instruments are available \citep[e.g.][]{1998Msngr..94....1A}. The total emission at these wavelengths is probably dominated by direct light from the stellar halo, but it is possible that the signature of a very extended dust distribution might still be visible in polarised light. Detailed polarised radiative transfer simulations should be used to test this. 

\subsection{Extension to other galaxies}

An obvious extension of the present study is to apply this type of modelling to a larger set of edge-on spiral galaxies, in particular to galaxies without strong starbursts or active galactic nuclei. The required ingredients are the availability of high-quality and sensitive UV imaging, and a full SED that covers the entire UV--mm wavelength range. An obvious starting point would be the sample studied by \citet{2014ApJ...789..131H}, which was also the starting point for \citetalias{2015arXiv151103789S}. A potential problem is that this sample contains galaxies with a range of inclinations, whereas from a radiative transfer point of view, it is crucial that the galaxies to be modelled are as close to edge-on as possible and that they have sufficient spatial resolution. 

A set of galaxies that would be perfect for these goals is the combined HEROES--NHEMESES sample \citep{2012IAUS..284..128H, 2013A&A...556A..54V}. It contains 19 nearby edge-on spiral galaxies that have been observed extensively over the entire wavelength range, including deep far-infrared and submm imaging with {\it{Herschel}}. Unfortunately, only a fraction of these galaxies have deep FUV data from {\it{GALEX}}. As demonstrated by \citet{2014ApJ...789..131H}, the UVOT instrument onboard the {\it{Swift}} mission forms an interesting alternative, with the added advantage that {\it{Swift}} is still gathering data. A new possibility is the recently launched {\it{Astrosat}} observatory \citep{2006AdSpR..38.2989A}, a multi-wavelength satellite that also carries a UV instrument onboard \citep[UVIT,][]{2012SPIE.8443E..1NK, 2014Ap&SS.354..143H}.

\section{Conclusions}

\citetalias{2015arXiv151103789S} have recently argued for the presence of an extended distribution of dust in the edge-on spiral galaxy NGC\,3628 based on radiative transfer modelling of the observed surface brightness distribution in the FUV band. To further investigate this claim, we have set up a panchromatic radiative transfer model for NGC\,3628, with the distribution of dust and young stellar populations from \citetalias{2015arXiv151103789S} and the distribution of evolved stellar populations from \citet{2015ApJS..219....4S} as the only ingredients. The conclusions of this investigation are the following:
\begin{itemize}
\item While our model does not use any constraints in the infrared domain and no fine-tuning at all, it reproduces the dust emission in the infrared--mm wavelength range almost perfectly.  
\item Modelling the same system without the thick dust component results in a systematic underestimation of the observed SED. 
\item A possibly crucial caveat in our analysis is the geometrical simplicity and non-uniqueness of our model. In particular, our model contains a smooth and extended young stellar disc, whereas NGC\,3628 is known to host a central starburst and the WISE 22~$\upmu$m image shows a clear central concentration. A much more extended suite of detailed radiative transfer simulations is required before the extended UV emission can be firmly attributed to scattering off dust at high latitudes. 
\item UV polarisation measurements would be an interesting way to have independent information on the nature of the extended UV emission.
\item This investigation could easily be extended to a larger suite of edge-on galaxies, as long as high-quality UV imaging observations and an SED covering UV to submm wavelengths are available. In particular, galaxies without a strong nuclear starburst would facilitate the interpretation.
\end{itemize}

\begin{acknowledgements}
The authors gratefully acknowledge funding from the Belgian Science Policy Office (BELSPO) and the Flemish Fund for Scientific Research (FWO-Vlaanderen). The research was supported in part by the Interuniversity Attraction Poles Programme initiated by the Belgian Science Policy Office (AP P7/08 CHARM). We thank Kwang-Il Seon, Jong-Ho Shinn and the anonymous referee for comments and suggestions that improved the analysis and discussion in this paper. 
\newline
This publication makes use of data products from the Sloan Digital Sky Survey. Funding for the SDSS and SDSS-II has been provided by the Alfred P. Sloan Foundation, the Participating Institutions, the National Science Foundation, the U.S. Department of Energy, the National Aeronautics and Space Administration, the Japanese Monbukagakusho, the Max Planck Society, and the Higher Education Funding Council for England. The SDSS Web Site is \url{http://www.sdss.org/}. The SDSS is managed by the Astrophysical Research Consortium for the Participating Institutions. The Participating Institutions are the American Museum of Natural History, Astrophysical Institute Potsdam, University of Basel, University of Cambridge, Case Western Reserve University, University of Chicago, Drexel University, Fermilab, the Institute for Advanced Study, the Japan Participation Group, Johns Hopkins University, the Joint Institute for Nuclear Astrophysics, the Kavli Institute for Particle Astrophysics and Cosmology, the Korean Scientist Group, the Chinese Academy of Sciences (LAMOST), Los Alamos National Laboratory, the Max-Planck-Institute for Astronomy (MPIA), the Max-Planck-Institute for Astrophysics (MPA), New Mexico State University, Ohio State University, University of Pittsburgh, University of Portsmouth, Princeton University, the United States Naval Observatory, and the University of Washington. This publication makes use of data products from the Two Micron All Sky Survey, which is a joint project of the University of Massachusetts and the Infrared Processing and Analysis Center/California Institute of Technology, funded by the National Aeronautics and Space Administration and the National Science Foundation. This publication makes use of data products from the Wide-field Infrared Survey Explorer, which is a joint project of the University of California, Los Angeles, and the Jet Propulsion Laboratory/California Institute of Technology, funded by the National Aeronautics and Space Administration.
\newline
This research has made use of the NASA/IPAC Extragalactic Database (NED) which is operated by the Jet Propulsion Laboratory, California Institute of Technology, under contract with the National Aeronautics and Space Administration. This research has made use of NASA's Astrophysics Data System bibliographic services. This research made use of Astropy, a community-developed core Python package for Astronomy \citep{2013A&A...558A..33A}.
\end{acknowledgements}

\bibliographystyle{aa} 
\bibliography{References}

\end{document}